\documentstyle[preprint,aps,epsfig]{revtex}
\tighten
\begin{document}
 \preprint{Preprint Numbers: \parbox[t]{45mm}{ANL-PHY-8897-TH-98\\
                                             MPG-VT-UR 124/98\\
}}

\title{Analysis of chiral and thermal susceptibilities}

\author{D.~Blaschke\footnotemark[1], 
A.~H\"oll\footnotemark[1], 
C.D. Roberts\footnotemark[2]
and S.~Schmidt\footnotemark[1]\vspace*{0.2\baselineskip}}
\address{\footnotemark[1]Fachbereich Physik, Universit\"at Rostock,
D--18051 Rostock, Germany\\\vspace*{0.2\baselineskip}
\footnotemark[2]Physics Division, Bldg. 203, Argonne National
Laboratory, Argonne IL 60439-4843, USA }
\date{Pacs Numbers: 11.10.Wx, 12.38.Mh, 24.85.+p, 05.70.Fh }

\maketitle

\begin{abstract}
We calculate the chiral and thermal susceptibilities for two confining
Dyson-Schwinger equation models of QCD with two light flavours, a
quantitative analysis of which yields the critical exponents, $\beta$ and
$\delta$, that characterise the second-order chiral symmetry restoration
transition.  The method itself is of interest, minimising the influence of
numerical noise in the calculation of the order parameter for chiral symmetry
breaking near the critical temperature.  For the more realistic of the two
models we find: $T_c \approx 153\,$MeV, and the non-mean-field values: $\beta
= 0.46 \pm 0.04$, $\delta = 4.3\pm 0.3$ and $1/(\beta \delta)= 0.54 \pm
0.05$, which we discuss in comparison with the results of other models.
\end{abstract}
\section{Introduction}
Phase transitions are characterised by the behaviour of an order parameter,
$\langle X \rangle$, the expectation value of an operator.  In the ordered
phase of a system: $\langle X \rangle \neq 0$, whereas in the disordered
phase $\langle X \rangle = 0$.  A phase transition is first-order if $\langle
X \rangle \to 0$ discontinuously, whereas it is second-order if $\langle X
\rangle \to 0$ continuously.  For a second-order transition, the length-scale
associated with correlations in the system diverges as $\langle X \rangle \to
0$ and one can define a range of critical exponents that characterise the
behaviour of certain macroscopic properties at the transition point.  For
example, in a system that is ferromagnetic for temperatures less than some
critical value, $T_c$, the magnetisation, $M$, in the absence of an external
magnetic field, behaves as $M \propto (T_c-T)^\beta$ for $T\sim T_c^-$, where
$\beta$ is the critical exponent.  At the critical temperature the behaviour
of the magnetisation in the presence of an external field, $h\to 0^+$,
defines another critical exponent, $\delta$: $M \propto h^{(1/\delta)}$.  In
a system that can be described by mean field theory these critical exponents
are
\begin{eqnarray}
\beta^{\rm MF}= 0.5\,,\; & & \delta^{\rm MF} = 3.0\,.
\end{eqnarray}

Equilibrium, second-order phase transitions can be analysed using the
renormalisation group, which leads to scaling laws that reduce the number of
independent critical exponents to just two: $\beta$ and
$\delta$~\cite{cpbook}.  It is widely conjectured that the values of these
exponents are fully determined by the dimension of space and the nature of
the order parameter.  This is the notion of {\it universality}$\,$; i.e.,
that the critical exponents are {\it independent} of a theory's microscopic
details and hence all theories can be grouped into a much smaller number of
universality classes according to the values of their critical exponents.  If
this is the case, the behaviour of a complicated theory near criticality is
completely determined by the behaviour of a simpler theory in the same
universality class.  So, when presented with an apparently complicated
theory, the problem is reduced to only that of establishing its universality
class.

Quantum chromodynamics is an asymptotically free theory; i.e., there is an
intrinsic, renormalisation-induced mass-scale, $\Lambda_{\rm QCD}$, and for
squared momentum transfer $Q^2 \gg \Lambda_{\rm QCD}$, the interactions
between quarks and gluons are weaker than Coulombic: $\alpha_{\rm S}(Q^2)\to
0$ as $Q^2 \to \infty$.  The study of QCD at finite temperature and baryon
number density proceeds via the introduction of the intensive variables:
temperature, $T$; and quark chemical potential, $\mu$.  These are additional
mass-scales, with which the coupling can {\it run} and hence, for $T\gg
\Lambda_{\rm QCD}$ and/or $\mu\gg \Lambda_{\rm QCD}$, $\alpha_{\rm
S}(Q^2=0,T,\mu)\sim 0$.  It follows that, at finite temperature and/or baryon
number density, there is a phase of QCD in which quarks and gluons are weakly
interacting, {\em irrespective} of the momentum transfer~\cite{collinsperry};
i.e., a quark-gluon plasma phase.  Such a phase of matter existed
approximately one microsecond after the big-bang.

At $T,\mu = 0$ the strong interaction is characterised by confinement and
dynamical chiral symmetry breaking (DCSB), effects which are tied to the
behaviour of $\alpha_{\rm S}(Q^2)$ at small-$Q^2$; i.e., its long-range
behaviour.  In a phase of QCD in which the coupling is uniformly small for
all $Q^2$, these effects are absent and the nature of the strong interaction
spectrum is qualitatively different.

The path followed in the transition to the plasma is also important because
it determines some observational consequences of the plasma's existence.  For
example~\cite{krishna}, the time-scale for the expansion of the early
universe: $\sim 10^{-5}\, {\rm s}$, is large compared with the natural
time-scale in QCD: $1/\Lambda_{\rm QCD} \sim 1\,{\rm fm}/c \sim
10^{-23}\,{\rm s}$, hence thermal equilibrium is maintained throughout the
QCD transition.  Therefore if the transition is second-order the ratio $B
:=\,$baryon-number/entropy, remains unchanged from that value attained at an
earlier stage in the universe's evolution.  However, a first-order transition
would be accompanied by a large increase in entropy density and therefore a
reduction in $B$ after the transition.  Hence the order of the QCD transition
constrains the mechanism for baryon number generation in models describing
the formation of the universe, since with a second-order transition this
mechanism is only required to produce the presently observed value of $B$ and
need not allow for dilution.  In the absence of quarks, QCD has a first-order
deconfinement transition, while with three or four massless quarks a
first-order chiral symmetry restoration transition is
expected~\cite{krishna}.  What of the realistic case with two light quark
flavours?

Based on the global chiral symmetry of QCD with two light quark flavours, it
has been argued~\cite{krishna} that this theory and the $N=4$ Heisenberg
magnet are in the same universality class.  As a field theory, the $N=4$
Heisenberg magnet is characterised by an interaction of the form
\begin{equation}
\sum_{i=1}^4\, \left\{ \case{1}{2} \mu^2 \phi_i^2(x) 
        + \case{1}{4} \lambda^4 \phi_i^4(x) \right\}\,,
\end{equation}
where $\mu^2$ is a function of temperature: $\mu^2\geq 0$ at or above the
critical temperature, $T_c^H$, but $\mu^2 <0$ for $T<T_c^H$.  If the
interaction strength, $\lambda$, depends smoothly on $T$ and remains positive
then, for $T<T_c^H$, the classical minimum of this potential is at
\begin{equation}
\phi_{\rm cl}^2 = \frac{-\mu^2}{\lambda} > 0\,.
\end{equation}
This model is familiar as the nonlinear $\sigma$-model, often used to
describe low-energy phenomena in QCD.  It has been explored thoroughly and
has a second order phase transition with critical exponents~\cite{neqfour}
\begin{eqnarray}
\beta^H= 0.38 \pm 0.01\,,\; & & \delta^H = 4.82 \pm 0.05\,.
\end{eqnarray}

One can examine the hypothesis that this model and QCD with two light quark
flavours are in the same universality class via numerical simulations.  Such
studies on an $8^3\times 4$ lattice suggest a second-order chiral phase
transition with critical exponents~\cite{kl94}
\begin{eqnarray}
\beta^{\rm lat}= 0.30 \pm 0.08\,,\; & & \delta^{\rm lat} = 4.3 \pm 0.5
\end{eqnarray}
but do not decide the question.\footnote{A review~\cite{el98} of results from
more recent simulations on larger lattices with lighter quarks reports a
significant dependence of these critical exponents on the lattice volume but
with their product approximately constant.  A value of $\delta\approx 1$ is
obtained, which is characteristic of a first-order transition.  These
unexpected results might be artefacts of finite lattice spacing because
introducing light dynamical quarks drives the simulations to stronger
coupling and hence coarser lattices.}  These results were obtained through an
analysis of the chiral and thermal susceptibilities; a technique that can be
applied in the study of any theory.  Herein we illustrate the method via an
analysis of two Dyson-Schwinger equation (DSE) models of QCD, which also
allows us to explore the hypothesis further.

Dyson-Schwinger equations provide a renormalisable, nonperturbative,
continuum framework for the exploration of strong interaction effects and
have been used extensively at $T=0$~\cite{rw94} in the study of confinement
and DCSB, and in the calculation of hadron
observables~\cite{tandy,pichowsky,ivanov}.  They have
recently~\cite{prl,thermo} found successful application at $T\neq 0$ and it
is these two models that we employ as exemplars herein.  In Sec.~II we
describe the models and in Sec.~III the analysis of their chiral and thermal
susceptibilities, and the evaluation of the associated critical exponents.
We summarise and conclude in Sec.~IV.

\section{Two Models}
Using a Euclidean metric, with $\{\gamma_\mu,\gamma_\nu\}=2\,\delta_{\mu\nu}$
and $\gamma_\mu^\dagger = \gamma_\mu$, the renormalised dressed-quark
propagator at $T\neq 0$ takes the form
\begin{equation}
S(p_{\omega_k}) := -i\vec{\gamma}\cdot \vec{p}\,\sigma_A(p_{\omega_k}) -
i\gamma_4 \omega_k \,\sigma_C(p_{\omega_k}) + \sigma_B(p_{\omega_k})\,,
\end{equation}
where $(p_{\omega_k}):= (\vec{p},\omega_k)$ with $\omega_k= (2 k + 1)\,\pi T$
the fermion Matsubara frequency, and $\sigma_{\cal F}(p_{\omega_k})$, ${\cal
F}=A,B,C$, are functions only of $|\vec{p}|^2$ and $\omega_k^2$.  The
propagator is obtained as a solution of the quark DSE
\begin{eqnarray}
S^{-1}(p_{\omega_k}) & :=& i\vec{\gamma}\cdot \vec{p} \,A(p_{\omega_k})
+ i\gamma_4\,\omega_k \,C(p_{\omega_k} ) + B(p_{\omega_k} )\\ 
&= &Z_2^A
\,i\vec{\gamma}\cdot \vec{p} + Z_2 \, (i\gamma_4\,\omega_k + m_{\rm bm})\, +
\Sigma^\prime(p_{\omega_k} ),
\label{qDSE} 
\end{eqnarray}
$m_{\rm bm}$ is the Lagrangian current-quark bare mass and the regularised
self energy is
\begin{eqnarray}
\Sigma^\prime(p_{\omega_k}) &=& i\vec{\gamma}\cdot
\vec{p}\,\Sigma_A^\prime(p_{\omega_k} ) 
+ i\gamma_4\,\omega_k\,\Sigma_C^\prime(p_{\omega_k} ) + 
\Sigma_B^\prime(p_{\omega_k})\,, 
\end{eqnarray}
with
\begin{eqnarray}
\Sigma_{\cal F}^\prime(p_{\omega_k}) =\int_{l,q}^{\bar\Lambda}\,
\case{4}{3}\,g^2\,D_{\mu\nu}(\vec{p}-\vec{q},\omega_k-\omega_l)
\,\case{1}{4}{\rm tr}\left[{\cal P}_{\cal F} \gamma_\mu
S(q_{\omega_l})\Gamma_\nu(q_{\omega_l};p_{\omega_k})\right]\,,
\label{regself}
\end{eqnarray}
where: ${\cal F}=A,B,C$; ${\cal P}_A:= -(Z_1^A/|\vec{p}|^2)i\vec{\gamma}\cdot
\vec{p}$, ${\cal P}_B:= Z_1 $, ${\cal P}_C:= -(Z_1/\omega_k)i\gamma_4$; and
$\int_{l,q}^{\bar\Lambda}:=\, T
\,\sum_{l=-\infty}^\infty\,\int^{\bar\Lambda}d^3q/(2\pi)^3$, with
$\int^{\bar\Lambda}$ a mnemonic to represent a translationally invariant
regularisation of the integral and $\bar\Lambda$ the regularisation
mass-scale.  In Eq.~(\ref{regself}), $\Gamma_\nu(q_{\omega_l};p_{\omega_k})$
is the renormalised dressed-quark-gluon vertex and $
D_{\mu\nu}(\vec{p},\Omega_k)$ is the renormalised dressed-gluon propagator.
($\Omega_k = 2 k\, \pi T$ is the boson Matsubara frequency.)

In renormalising the quark DSE we require that
\begin{equation}
\label{subren}
\left.S^{-1}(p_{\omega_0})\right|_{|\vec{p}|^2+\omega_0^2=\zeta^2} = 
        i\vec{\gamma}\cdot \vec{p} + i\gamma_4\,\omega_0 + m_R\;,
\end{equation}
which entails that the renormalisation constants are
\begin{eqnarray}
Z_2^A(\zeta,\bar\Lambda) 
& = & 1- \Sigma_A^\prime(\zeta^-_{\omega_0};{\bar\Lambda}),\\
Z_2(\zeta,\bar\Lambda) 
& = & 1- \Sigma_C^\prime(\zeta^-_{\omega_0};{\bar\Lambda}),\\
m_R(\zeta) & = & Z_2 m_{\rm bm}({\bar\Lambda}^2) +
\Sigma_B^\prime(\zeta^-_{\omega_0};{\bar\Lambda}),
\end{eqnarray}
where $(\zeta^-_{\omega_0})^2 := \zeta^2 - \omega_0^2$, and the renormalised
self energies are
\begin{equation}
\begin{array}{rcl}
{\cal F}(p_{\omega_k};\zeta) & = & 
\xi_{\cal F} + \Sigma_{\cal F}^\prime(p_{\omega_k};{\bar\Lambda})
    - \Sigma_{\cal F}^\prime(\zeta^-_{\omega_0};{\bar\Lambda})\,,
\end{array}
\end{equation}
${\cal F}=A,B,C$, $\xi_A = 1 = \xi_C$ and $\xi_B=m_R(\zeta)$.

So far no approximations or truncations have been made but to continue we
must know the form of $\Gamma_\nu(q_{\omega_l};p_{\omega_k})$ and
$D_{\mu\nu}(\vec{p},\Omega_k)$ in Eq.~(\ref{regself}).  These Schwinger
functions satisfy DSEs.  However, the study of those equations is rudimentary
even at $T=0$ and there are no studies for $T\neq 0$.  To proceed we use the
$T=0$ results as a qualitative guide and employ exploratory {\it Ans\"atze}
for $\Gamma_\nu(q_{\omega_l};p_{\omega_k})$ and
$D_{\mu\nu}(\vec{p},\Omega_k)$.  This is where model parameters enter.

The structure of the dressed fermion-gauge-boson vertex has been much
considered~\cite{ayse97}.  As a connected, irreducible three-point function
it should be free of light-cone singularities in covariant gauges; i.e., it
should be regular at $(\vec{p}-\vec{q})^2 + (\omega_k-\omega_l)^2=0$.  A
range of {\it Ans\"atze} with this property have been proposed and
employed~\cite{hawes94} and it has become clear that the judicious use of the
rainbow truncation
\begin{equation}
\label{rainbow}
\Gamma_\nu(q_{\omega_l};p_{\omega_k}) = \gamma_\nu
\end{equation}
in Landau gauge provides reliable results~\cite{mr97}.  This is the {\it
Ansatz} employed in Refs.~\cite{prl,thermo} and we use it herein.  With this
truncation a mutually consistent constraint is $Z_1 = Z_2$ and $Z_1^A =
Z_2^A$~\cite{mr97}.

With $\Gamma_\nu(q_{\omega_l};p_{\omega_k})$ regular, the analytic properties
of the kernel in the quark DSE are determined by those of
$D_{\mu\nu}(p_{\Omega_k})$, which in Landau gauge has the general form
\begin{equation}
g^2 D_{\mu\nu}(p_{\Omega_k}) = 
P_{\mu\nu}^L(p_{\Omega_k} ) \Delta_F(p_{\Omega_k} ) + 
P_{\mu\nu}^T(p_{\Omega_k}) \Delta_G(p_{\Omega_k}  ) \,,
\end{equation}
\begin{eqnarray}
P_{\mu\nu}^T(p_{\Omega_k}) & \equiv &\left\{
\begin{array}{c}
0; \; \mu\;{\rm and/or} \;\nu = 4,\\
\displaystyle
\delta_{ij} - \frac{p_i p_j}{p^2}; \; \mu,\nu=i,j\,=1,2,3\;,
\end{array}\right.
\end{eqnarray}
with $P_{\mu\nu}^T(p_{\Omega_k}) + P_{\mu\nu}^L(p_{\Omega_k}) =
\delta_{\mu\nu}- p_\mu p_\nu/{\sum_{\alpha=1}^4 \,p_\alpha p_\alpha}$;
$\mu,\nu= 1,\ldots, 4$.  A ``Debye-mass'' for the gluon appears as a
$T$-dependent contribution to $\Delta_F$.  Considering $D_{\mu\nu}(k)$ at
$T=0$, a perturbative analysis at two-loop order provides a quantitatively
reliable estimate for $k^2 > 1$-$2\,$GeV$^2$, with higher order terms
providing corrections of only $\sim 10$\%.  However, for $k^2<1\,$GeV$^2$
nonperturbative methods are necessary.  Studies of the gluon DSE in axial
gauge~\cite{atkinson}, where ghost contributions are absent, or in Landau
gauge~\cite{pennington}, when their contributions are small, indicate that
$D_{\mu\nu}(k)$ is significantly enhanced in the vicinity of $k^2 = 0$
relative to a free gauge-boson propagator, and that the enhancement persists
to $k^2 \sim 1\,$GeV$^2$.  Due to the truncations involved these studies are
not quantitatively reliable but this behaviour has been modelled successfully
as a distribution located in the vicinity of $k^2 = 0$~\cite{mr97,fr}.

\subsection{Infrared-dominant Model}
A particularly simple and illustratively useful model is obtained with
\begin{equation}
\label{dmn}
\Delta_F(p_{\Omega_k}) = \Delta_G(p_{\Omega_k}) 
= 2 \pi^3 \,\frac{\eta^2}{T}\, \delta_{k0}\, \delta^3(\vec{p})\,,
\end{equation}
which is a generalisation to $T\neq 0$ of the model introduced in
Ref.~\cite{mn83}, where $\eta\approx 1.06\,$GeV is a mass-scale parameter
fixed by fitting $\pi$- and $\rho$-meson masses.  As an infrared-dominant
model Eq.~(\ref{dmn}) does not represent well the behaviour
$D_{\mu\nu}(p_{\Omega_k})$ away from $p_{\Omega_k}^2 \simeq 0$, and hence
there are some model-dependent artefacts.  However, these artefacts are
easily identified and, because of its simplicity, the model has provided a
useful means of elucidating many of the qualitative features of more
sophisticated {\it Ans\"atze}.

Using Eqs.~(\ref{rainbow}) and (\ref{dmn}) the quark DSE is
ultraviolet-finite, the cutoff can be removed and the renormalisation point
taken to infinity, so that Eq.~(\ref{qDSE}) becomes the algebraic
equations
\begin{eqnarray}
\label{beqnfour}
\eta^2 m^2 & = & B^4 + m\, B^3 + \left(4 p_{\omega_k}^2 - \eta^2 -
        m^2\right) B^2 -m\,\left( 2\,{{\eta }^2} + {m^2} +
        4\, p_{\omega_k}^2 \right)B   \,,     \\ 
A(p_{\omega_k}) & = & C(p_{\omega_k}) = 
        \frac{2 B(p_{\omega_k})}{m +B(p_{\omega_k})}\,,
\end{eqnarray}
with $Z_2^A= 1 = Z_2$ and $m=m_R=m_{\rm bm}$: $m=0$ defines the chiral limit.
This DSE-model of QCD has coincident, second-order deconfinement and chiral
symmetry restoration phase transitions at a critical temperature $T_c^{\rm
IR}\approx 0.16\,\eta$~\cite{thermo}.

\subsection{Ultraviolet-improved Model}
\label{secuvi}
An improvement over Eq.~(\ref{dmn}) is obtained by correcting the
large-$p_{\Omega_k}^2$ behaviour so as to better represent the interaction at
short-distances.  The one-parameter model
\begin{eqnarray}
\label{uvpropf}
\Delta_F(p_{\Omega_k}) & = & {\cal D}(p_{\Omega_k};m_D)\,,\\
\label{uvpropg}
\Delta_G(p_{\Omega_k}) & = & {\cal D}(p_{\Omega_k};0)\,,\\
\label{delta}
 {\cal D}(p_{\Omega_k};m) & := & \case{16}{9}\,\pi^2 \, \left[ 
\frac{2\pi}{T} m_t^2 \delta_{0\,k} \delta^3(\vec{p}) + 
\frac{1-{\rm e}^{
\left[-\right(|\vec{p}|^2+\Omega_k^2+ m^2 \left)/(4m_t^2)\right]}}
        {|\vec{p}|^2+\Omega_k^2+ m^2} \right]\, ,
\end{eqnarray}
where $m_{\rm D}^2 = (8/3)\, \pi^2 T^2$ is the perturbatively evaluated
``Debye-mass''\footnote{The influence of the Debye-mass on finite-$T$
observables is qualitatively unimportant, even in the vicinity of the chiral
symmetry restoration transition.  The ratio of the coefficients in the two
terms in Eq.~(\protect\ref{delta}) is such that the long-range effects
associated with $\delta_{0\,k} \delta^3(p)$ are completely cancelled at
short-distances; i.e., for $|\vec{x}|^2\,m_t^2\ll 1$.}, achieves this.  This
gluon propagator provides a generalisation to $T\neq 0$ of the model explored
in Ref.~\cite{fr} where the parameter $m_t$ is a mass-scale that marks the
boundary between the perturbative and nonperturbative domains.  The value
$m_t=0.69\,{\rm GeV}=1/0.29\,{\rm fm}$ is fixed by requiring a good
description of a range of $\pi$- and $\rho$-meson properties.  In this case
the DSE yields a pair of coupled, nonlinear integral equations that must be
solved subject to the renormalisation boundary conditions, and $m_R=0$
defines the chiral limit.  This model also has coincident, second-order
deconfinement and chiral symmetry restoration transitions, with the critical
temperature $T_c^{\rm UV}\approx 0.15\,$GeV~\cite{prl}.

\section{Chiral and Thermal Susceptibilities}
In the study of dynamical chiral symmetry breaking an order parameter often
used is the quark condensate, $\langle \bar q q\rangle_\zeta$.  In QCD in the
chiral limit this order parameter is defined via
the quark propagator~\cite{mr97}:
\begin{equation}
\label{qbarq}
-\langle \bar q q\rangle_\zeta:= N_c\,
         \lim_{\bar\Lambda\to \infty}
        Z_4(\zeta,\bar\Lambda)\,
        \int_{l,q}^{\bar\Lambda}
        \frac{B_0(p_{\omega_l})}
        {|\vec{p}|^2 A_0(p_{\omega_l})^2
        + \omega_l^2 C_0(p_{\omega_l})^2 
        + B_0(p_{\omega_l})^2}\,,
\end{equation}
for each massless quark flavour, where the subscript ``$0$'' denotes that the
scalar functions: $A_0$, $B_0$, $C_0$, are obtained as solutions of
Eq.~(\ref{qDSE}) in the chiral limit, and $Z_4(\zeta,\bar\Lambda)$ is the
mass renormalisation constant: $Z_4(\zeta,\bar\Lambda) \,m_R(\zeta) =
Z_2(\zeta,\bar\Lambda)\, m_{\rm bm}(\bar\Lambda)$.  The functions have an
implicit $\zeta$-dependence.  From Eq.~(\ref{qbarq}) it is clear that an
equivalent order parameter for the chiral transition is
\begin{equation}
{\cal X} := B_0(\vec{p}=0,\omega_0)\,,
\end{equation}
which was used in Refs.~\cite{prl,thermo}.  Thus the zeroth Matsubara mode
determines the character of the chiral phase transition, a conjecture
explored in Ref.~\cite{jackson96}.

To accurately characterise the chiral symmetry restoration transitions in the
two models introduced above, we examine closely the chiral and thermal
susceptibilities and their scaling behaviour near the critical point.  This
allows a determination of the critical temperature, $T_c$, and exponents
$\beta$ and $\delta$, as we explain in the appendix.  In the notation of the
appendix, the ``magnetisation'' is
\begin{equation}
M(t,h):= B(\vec{p}=0,\omega_0)\,,
\end{equation}
i.e., the value in the infrared of the scalar piece of the quark self energy
obtained as the $m_R$- and $T$-dependent solution of Eq.~(\ref{qDSE}).

\subsection{Critical Exponents of the Infrared-dominant Model}
\label{secir}
In the chiral limit, Eq.~(\ref{beqnfour}) has the Nambu-Goldstone mode
solution
\begin{eqnarray}
\label{ngsoln}
B(p_{\omega_k}) & = &\left\{
\begin{array}{lcl}
\sqrt{\eta^2 - 4  p_{\omega_k}^2}\,, 
        & & p_{\omega_k}^2<\case{\eta^2}{4}\\
0\,, & & {\rm otherwise}
\end{array}\right.\\
C(p_{\omega_k}) & = &\left\{
\begin{array}{lcl}
2\,, & & p_{\omega_k}^2<\case{\eta^2}{4}\\
\case{1}{2}\left( 1 + \sqrt{1 + \case{2 \eta^2}{p_{\omega_k}^2}}\right)
\,,& & {\rm otherwise}\,,
\end{array}\right.
\end{eqnarray}
and hence
\begin{equation}
\label{MIR}
M(t,0)= 2 \pi\,\left(\frac{\eta}{2\pi} + T\right)^{\case{1}{2}}
                \left(\frac{\eta}{2\pi} - T\right)^{\case{1}{2}}\,.
\end{equation}
From Eq.~(\ref{MIR}) we read that
\begin{equation}
\label{mnexa}
T_c^{\rm IR}= \frac{\eta}{2\pi}\approx 0.159155\,\eta\,,\; 
        \beta^{\rm IR}= \frac{1}{2}\,.
\end{equation}
To determine $\delta$ we use Eq.~(\ref{beqnfour}) at $T=T_c$ to obtain
\begin{equation}
\eta^2 m^2 = M(0,h)^4 + m\, M(0,h)^3  + 
        m^2\, M(0,h)^2 - m (3\eta^2 + m^2)\, M(0,h) 
\end{equation}
and suppose that, for $m\sim 0$, $M(0,h) = a \,m^{1/\delta}$.  Consistency
requires 
\begin{equation}
\label{mnexb}
\delta^{\rm IR} = 3\,.
\end{equation}
That the chiral symmetry restoration transition in this model is
characterised by mean field critical exponents is not surprising because the
interaction described by Eq.~(\ref{dmn}) is a constant in configuration
space.  Mean field critical exponents are also obtained in chiral random
matrix models of QCD~\cite{jackson96,wettig97}.

To illustrate the evaluation of the critical temperature and exponents using
the chiral and thermal susceptibilities we use Eqs.~(\ref{beqnfour}),
(\ref{defchih}) and (\ref{defchit}) to obtain
\begin{eqnarray}
\label{chihmn}
\chi_h(T,h) & = & - \frac{2\,M(T,h)\,T\,b_-\,[1-M(T,h)\,b_-]- M(T,h)\,T\,b_+}
        {2\,M(T,h)\,b_-\,b_+\,[1-M(T,h)\,b_-] - b_-\,b_+ - M(T,h)\,b_-}\,,\\
\label{chiTmn}
\chi_T(T,h)  &= &\frac{8\pi^2\,T\,M(T,h)^2\,b_-^2 - 2\,M(T,h)\,b_-\,b_+
                        h\,[1-M(T,h)\,b_-]- M(T,h)\,h\,b_+^2}
        {2\,M(T,h)\,b_-\,b_+\,[1-M(T,h)\,b_-] - b_-\,b_+ - M(T,h)\,b_-}\,,
\end{eqnarray}
where $b_\pm := M(T,h) \pm h T$.  In Fig.~\ref{mncs} we plot the chiral
susceptibility.  The temperature dependence is typical of this quantity, with
the peak increasing in height and becoming narrower as $h\to 0^+$; i.e., as
the external source for chiral symmetry breaking is removed.  To understand
this behaviour, recall that the chiral susceptibility is the derivative of
the order parameter with-respect-to the explicit chiral symmetry breaking
mass.  Denote the typical mass-scale associated with DCSB by $M_\chi$.  For
$h \gg M_\chi$, explicit chiral symmetry breaking dominates, with the order
parameter ${\cal X}\sim h$ and insensitive to $T$, and hence $\chi_h
\approx\,$const.  For $h \sim M_\chi$, ${\cal X}$ begins to vary with $T$
because the origin of its magnitude changes from the explicit mass to the
DCSB mechanism as $T$ passes through the pseudocritical temperature, $T_{\rm
pc}^h$.  This is reflected in $\chi_h$ as the appearance of a peak at $T_{\rm
pc}^h$.  For $h\ll M_\chi$, ${\cal X}\sim h$ until very near $T_{\rm pc}^h$
when the scale of DCSB overwhelms $h$ and ${\cal X}\sim M_\chi$.  The change
in ${\cal X}$ is rapid leading to the behaviour observed in $\chi_h$.  The
thermal susceptibility is plotted in Fig.~\ref{mnts} and has qualitatively
similar features.

In Table~\ref{taba} we present the pseudocritical points and peak heights
obtained for $h$ in the scaling window, defined as the domain of $h$ for
which
\begin{equation}
\frac{t_{\rm pc}^h}{t_{\rm pc}^t} = {\rm const.}\,;
\end{equation}
i.e., the values of $h$ for which Eqs.~(\ref{pch}) and (\ref{pct}) are valid.
Based on Eqs.~(\ref{deltaslope}) and (\ref{betaslope}), using the tabulated
values, one obtains $z_h^{\rm IR}$ and $z_t^{\rm IR}$ from linear fits to the
curves: $\log \chi_h^{\rm pc}$-versus-$\log h$ and $\log \chi_T^{\rm
pc}$-versus-$\log h$, respectively.  This yields
\begin{equation}
\label{zedsir}
z_h^{\rm IR} = 0.666,\; z_t^{\rm IR} = 0.335\,,
\end{equation}
and hence $\beta_\chi^{\rm IR} = 0.499$ and $\delta_\chi^{\rm IR} = 2.99$, as
listed in Table~\ref{tabc}.  These values are in excellent agreement with the
exact (mean field) results, Eqs.~(\ref{mnexa}) and (\ref{mnexb}).  With the
value of
\begin{equation}
\frac{1}{(\beta\delta)^{\rm IR}} = 1 - z_h^{\rm IR} + z_t^{\rm IR}  = 0.670\,,
\end{equation}
$T_c^{\rm IR}$ can be obtained in a variational procedure based on
Eq.~(\ref{pch}): it is that value which minimises the standard deviation
between $\log(T^h_{\rm pc}-T_c^{\rm IR}) - 1/(\beta\delta)^{\rm IR}\,\log h$
and a constant.  This yields $ T_c^{\rm IR} = 0.159155\,\eta$ again in
excellent agreement with Eq.~(\ref{mnexa}).  The value in Table~\ref{tabc} is
obtained with $\eta=1.06\,$GeV~\cite{mn83}.  Applying the same procedure to
\mbox{$\log(T^T_{\rm pc}-T_c^{\rm IR}) - 1/(\beta\delta)^{\rm IR}\,\log h$},
yields $T_c^{\rm IR} = 0.159151\,\eta$.

\subsection{Critical Exponents of the Ultraviolet-improved Model}
In this case the quark DSE must be solved numerically, as in
Refs.~\cite{prl,fr}.  In these calculations we used a $3$-momentum grid with
$96$ points and we renormalised at $\zeta=9.47\,$GeV, the value at which the
parameter $m_t(=0.69\,$GeV$)$ was fixed~\cite{fr}.  The chiral and thermal
susceptibilities for a range of values of $h$ are plotted in Figs.~\ref{frcs}
and \ref{frts}, and the pseudocritical points and peak heights obtained for
values of $h$ in the scaling window are presented in Table~\ref{tabb}.

As observed in Sec.~\ref{secir}, one obtains $z_h^{\rm UV}$ and $z_t^{\rm
UV}$ from linear fits to the curves $\log \chi_h^{\rm pc}$-versus-$\log h$
and $\log \chi_T^{\rm pc}$-versus-$\log h$, respectively.  The data and fits
are presented in Fig.~\ref{figchis} and yield
\begin{equation}
\label{zeds}
z_h^{\rm UV} = 0.77 \pm 0.02\,,\; z_t^{\rm UV} = 0.28 \pm 0.04\,,
\end{equation}
with the corresponding results for $\beta$ and $\delta$ listed in the first
column of Table~\ref{tabc}.\footnote{Our quoted error bounds the slope of the
linear fit.  It is calculated from the slope of linear fits to the two
endpoint values when they are displaced vertically, in opposite directions,
by the standard deviation of the fit to all the tabulated results.}  For this
model only, as a check and demonstration of consistency, the values of
$T_c^{\rm UV}$ and $1/(\beta\delta)^{\rm UV}$ were calculated using a
variational procedure based on Eqs.~(\ref{pch}) and (\ref{pct}): the values
of $T_c^{\rm UV}$ and $1/(\beta\delta)^{\rm UV}$ were varied in order to
minimise the standard deviation in a linear fit to \mbox{$\log(T_{\rm
pc}-T_c^{\rm UV}) - 1/(\beta\delta)^{\rm UV}\,\log h$}.  The difference
between using $T_{\rm pc}^h$ and $T_{\rm pc}^T$ is less than the error quoted
in the table.

In Ref.~\cite{prl} the values of $\beta$ and $T_c$ in the
ultraviolet-improved model were calculated directly from the magnetisation
order parameter; i.e., using Eq.~(\ref{Mtzero}), with the results $\beta =
0.33 \pm 0.3$ and $T_c \approx 152\,$MeV.  There is a discrepancy in the
value of $\beta$.  We expect that the result obtained herein is more accurate
because our method avoids the numerical noise associated with establishing
the precise behaviour of the order parameter in the vicinity of the critical
temperature.

\section{Summary and Conclusions}
A primary purpose of this study was an illustration of the method by which
one can calculate the critical exponents that characterise a chiral symmetry
restoration transition, $\beta$ and $\delta$, using the chiral and thermal
susceptibilities.  For this purpose we chose two Dyson-Schwinger equation
models of two-light-flavour QCD that have been applied
successfully~\cite{prl,thermo,basti} in phenomenological studies of QCD at
finite temperature and density.  The method is reliable and should have a
wide range of application because it is more accurate in the presence of
numerical noise than a straightforward analysis of the chiral symmetry
(magnetisation) order parameter.

We established that our finite temperature extension of the infrared-dominant
model of Ref.~\cite{mn83} is characterised by mean field critical exponents,
listed in Table~\ref{tabc}.  It is therefore not in the universality class
expected~\cite{krishna,kl94} of two-light-flavour lattice-QCD.  However, the
critical temperature is consistent with that estimated in lattice
simulations.  This fits an emerging pattern that DSE models whose mass-scale
parameters are fixed by requiring a good description of hadron observables at
$T=0$, yield a reliable estimate of the critical temperature for chiral
symmetry restoration.  It is a quantity that is not too sensitive to details
of the model.

Consistent with this observation, the critical temperature in the ultraviolet
improved model of Ref.~\cite{fr} also agrees with that estimated in lattice
simulations.  The critical exponent $\delta$ agrees with the value obtained
for two-light-flavour lattice-QCD and is consistent with that of the $N=4$
Heisenberg magnet.  However, the difference between the value of $\beta$
obtained in the model and that in lattice simulations is significant.  It is
unlikely that numerical errors in our study are the cause of this
discrepancy.  The values of the critical exponents, and their product,
establish that this model is not mean field in character.  They also
establish that the ultraviolet-improved model, which provides a good
description of low-energy $\pi$- and $\rho$-observables, is not in the same
universality class as the $N=4$ Heisenberg magnet.

The difference between the infrared-dominant model and the
ultraviolet-improved one is the value of $m_t$; i.e., the mass scale that
marks the boundary between strong and weak coupling.  In the limit $m_t\to
\infty$, the infrared-dominant model is recovered from the
ultraviolet-improved one: in this limit the interaction is always strong.
Our results therefore demonstrate that the critical exponents are sensitive
to the particular manner in which the theory makes the transition from strong
to weak coupling.  This should be expected since that evolution is a
determining characteristic of the $\beta$-function of a renormalisable
theory, one which a chiral symmetry restoration transition must be sensitive
to.

The large-$p^2$ behaviour of the gluon propagator in the ultraviolet-improved
model, although better than that in the infrared-dominant model, is still
inadequate.  Its renormalisation group properties are more like those of
quenched-QED than QCD because of the absence of the logarithmic suppression
of the running coupling characteristic of asymptotically free theories.  This
is corrected in the model of Ref.~\cite{mr97}, which has more in common with
QCD at $T=0$ and whose finite temperature properties can therefore assist in
better understanding the details of the chiral symmetry restoration
transition in two-light-flavour QCD.

\acknowledgements D.B. and S.S. acknowledge the hospitality of the Physics
Division at Argonne National Laboratory, and C.D.R. that of the Department of
Physics at the University of Rostock during visits in which parts of this
work were conducted.  We are also grateful to the faculty and staff at
JINR-Dubna for their hospitality during the workshop on {\em Deconfinement at
Finite Temperature and Density} in October 1997.  This work was supported in
part by Deutscher Akademischer Austauschdienst; the US Department of Energy,
Nuclear Physics Division, under contract number W-31-109-ENG-38; the National
Science Foundation under grant no.~INT-9603385; and benefited from the
resources of the National Energy Research Scientific Computing Center.

\appendix
\section*{Critical Exponents from Susceptibilities}
Consider the free energy of a theory, represented by
\begin{equation}
f=f(t,h)\,,
\end{equation}
where $t:= T/T_c - 1$ is the reduced temperature and $h:= m/T$ is the
explicit source of chiral symmetry breaking measured in units of the
temperature; it is analogous to an external magnetic field.  Since
correlation lengths diverge in a second-order transition it follows that for
$t,h\to 0$ the free energy is a generalised homogeneous function; i.e.,
\begin{equation}
f(t,h) = \frac{1}{b}\,f(t \,b^{y_t},h \,b^{y_h}) \,.
\end{equation}
This entails the following behaviour of the ``magnetisation''
\begin{eqnarray}
M(t,h) & := &
\left.\frac{\partial\, f(t,h)}
        {\!\!\!\!\!\!\partial h}\right|_{t\;{\rm fixed}}\,,\\
M(t,h) & = & b^{y_h-1}\,M(t \,b^{y_t},h \,b^{y_h})\,.
\end{eqnarray}
The scaling parameter, $b$, is arbitrary and along the trajectory $|t|
b^{y_t}= 1$ one has
\begin{eqnarray}
M(t,h) & = &|t|^{(1-y_h)/y_t}\,M({\rm sgn}(t), h \,|t|^{-y_h/y_t})\,, \\
\label{Mtzero}
M(t,0) & \propto& |t|^\beta\,,\; \beta:= \frac{1-y_h}{y_t}\,.
\end{eqnarray}
Alternatively, along the trajectory $h b^{y_h}= 1$
\begin{eqnarray}
M(t,h) & = &h^{(1-y_h)/y_h}\,M(t \,h^{-y_t/y_h},1)\,, \\
M(0,h) & \propto& h^{1/\delta}\,,\; \delta:= \frac{y_h}{1-y_h}\,.
\end{eqnarray}
This defines the critical behaviour and provides that direct means of
extracting the critical exponent $\beta$ employed in Refs.~\cite{prl,thermo}.
However, because of numerical noise, it can be difficult to extract
quantitatively accurate results using this method.

The critical exponents can also be determined by studying the pseudocritical
behaviour of the chiral and thermal susceptibilities, defined respectively as
\begin{eqnarray}
\label{defchih}
\chi_h(t,h)& := &
\left.\frac{\partial\, M(t,h)}
        {\!\!\!\!\!\!\partial h}\right|_{t\;{\rm fixed}}\,,\\
\label{defchit}
\chi_t(t,h) &:= &
\left.\frac{\partial\, M(t,h)}
        {\!\!\!\!\!\!\partial t}\right|_{h\;{\rm fixed}}\,.
\end{eqnarray}
For convenience, we often use $\chi_T(T,h):= (1/T_c)\,\chi_t(t,h)$.

For $t,h\to 0^+$, along $h b^{y_h}= 1$, one has
\begin{eqnarray}
\label{chih}
\chi_h(t,h) &  = & h^{(1-2y_h)/y_h}\,\chi_h(t \,h^{-y_t/y_h},1)\,,\\
\label{chit}
\chi_t(t,h) &  = & h^{(1-y_h - y_t)/y_h}\,\chi_t(t \,h^{-y_t/y_h},1)\,.
\end{eqnarray}

At each $h$, $\chi_h(t,h)$ and $\chi_t(t,h)$ are smooth functions of $t$.
Suppose they have maxima at $t_{\rm pc}^h$ and $t_{\rm pc}^t$, respectively,
described as the pseudocritical points.  Consider the chiral susceptibility.
At its maximum
\begin{eqnarray}
0 & = & \left. \frac{\partial}{\partial t} \chi_h(t,h) \right|_{t_{\rm pc}^h}\\
  & = & \left. 
        h^{(1-2y_h)/y_h}\,\frac{\partial}{\partial t} 
        \left(t \,h^{-y_t/y_h}\right)\,
        \left[\frac{\partial}{\partial z}\chi_h(z,1)
        \right]_{z=t \,h^{-y_t/y_h}}\,\right|_{t_{\rm pc}^h}\,,
\end{eqnarray}
which entails that
\begin{equation}
\label{pch}
t_{\rm pc}^h = K_h\,  h^{y_t/y_h} = K_h\,  h^{1/(\beta \delta)}\,,
\end{equation}
where $K_h$ is an undetermined constant.  Similarly,
\begin{equation}
\label{pct}
t_{\rm pc}^t = K_t\,  h^{y_t/y_h} = K_t\,  h^{1/(\beta \delta)}\,.
\end{equation}
Since $\beta\delta >0$, it follows that the pseudocritical points approach
the critical point, $t=0$, as $h\to 0^+$.  It follows from Eqs.~(\ref{pch})
and (\ref{pct}) that at the pseudocritical points
\begin{eqnarray}
\label{deltaslope}
\chi_h^{\rm pc} & := & \chi_h(t_{\rm pc}^h,h) \propto h^{-z_h}\,,\;
        z_h:= 1 - \case{1}{\delta} \,,\\
\label{betaslope}
\chi_t^{\rm pc} & := & \chi_t(t_{\rm pc}^t,h) 
        \propto h^{-z_t}\,,
        \;z_t:= \case{1}{\beta\delta}\,(1-\beta)\,.
\end{eqnarray}

Thus by locating the pseudocritical points and plotting the peak-height of
the susceptibilities as a function of $h$ one can obtain values of $T_c$,
$\beta$ and $\delta$.

%

\begin{table}[h]
\begin{tabular}{ddddd}
$\log h$     & $T_{\rm pc}^h/\eta$  & $\chi_h^{\rm pc}/\eta$ 
        & $T_{\rm pc}^T/\eta$  & $\chi_T^{\rm pc}$ \\ \hline
-5.0 & 0.15921 & 707.0 & 0.15917 & 248.5   \\
-4.3 & 0.15931 & 241.9 & 0.15920 & 145.4   \\
-4.0 & 0.15941 & 152.9 & 0.15923 & 115.3   \\
-3.3 & 0.15990 &  52.19 & 0.15939 & 67.33   \\
-3.0 & 0.16034 &  32.91& 0.15953 &  53.34  \\
-2.3 & 0.16268 &  11.31& 0.16052 &  30.91  
\end{tabular}
\caption{The pseudocritical points and peak heights for the chiral and
thermal susceptibilities in the infrared-dominant model, obtained from
Eqs.~(\protect\ref{chihmn}) and (\protect\ref{chiTmn}) respectively.
\label{taba}}
\end{table}
\begin{table}[h]
\begin{tabular}{ccccc}
                  & IR Dominant & UV Improved & O$(4)$ & Lattice 
                        \\\hline 
$\delta$& 3.0~ & 4.3  $\pm$ 0.3 & 4.82 $\pm$ 0.05 & 4.3 $\pm $ 0.5 \\
$\beta$ & 0.50 & 0.46 $\pm$ 0.04   & 0.38 $\pm$ 0.01  & 0.30 $\pm$ 0.08  \\
$\frac{1}{\beta \delta}$ 
        & 0.67   &   0.54 $\pm$ 0.05 & 0.55 $\pm$ 0.02 & 0.77 $\pm $ 0.14 \\
$T_c\,$(MeV)  & 168.7 & 153.5 $\pm$ 0.1      & -- &  140 \ldots~160 
\end{tabular}
\caption{Critical exponents and temperature for the models considered herein
and a comparison with the results in the $N=4$ Heisenberg
magnet~\protect\cite{neqfour}, labelled as O$(4)$, and lattice simulations of
QCD with two light flavours~\protect\cite{kl94}.
\label{tabc}}
\end{table}
\begin{table}[h]
\begin{tabular}{ddddd}
$\log h$     & $T_{\rm pc}^h\,$(GeV)  & $\chi_h^{\rm pc}\,$(GeV)
        & $T_{\rm pc}^T\,$(GeV)  & $\chi_T^{\rm pc}$ \\ \hline
-4.30 & 0.15464 & 896.3 & 0.15379 & 67.72   \\
-4.00 & 0.15515 & 530.7 & 0.15394 & 55.70   \\
-3.70 & 0.15571 & 303.8 & 0.15422 & 45.70   \\
-3.52 & 0.15627 & 224.9 & 0.15443 &  40.65  \\
-3.40 & 0.15677 & 181.8 & 0.15460 &  37.37  \\
-3.30 & 0.15729 & 154.9 & 0.15487 &  35.00 \\
 -3.15 & 0.15795 & 120.3 & 0.15508 &  31.64 \\
 -3.04 & 0.15840 & 97.21 & 0.15534 &  29.32  \\
 -3.0  & 0.15872 &  90.03& 0.15536 &  28.39
\end{tabular}
\caption{The pseudocritical points and peak heights for the chiral and
thermal susceptibilities in the ultraviolet-improved model, obtained from the
numerical solution of Eq.~(\protect\ref{qDSE}) with the gluon propagator of
Eqs.~(\protect\ref{uvpropf})--(\protect\ref{delta}). From the dependence of
the peak heights on the number of points in the $\vec{p}$-array, we estimate
a systematic 1.5\% error in $\chi_h^{\rm pc}$ and 10\% in $\chi_T^{\rm pc}$.
\label{tabb}}
\end{table}
\setcounter{figure}{0}
\begin{figure}[h]
 \centering{\
 \epsfig{figure=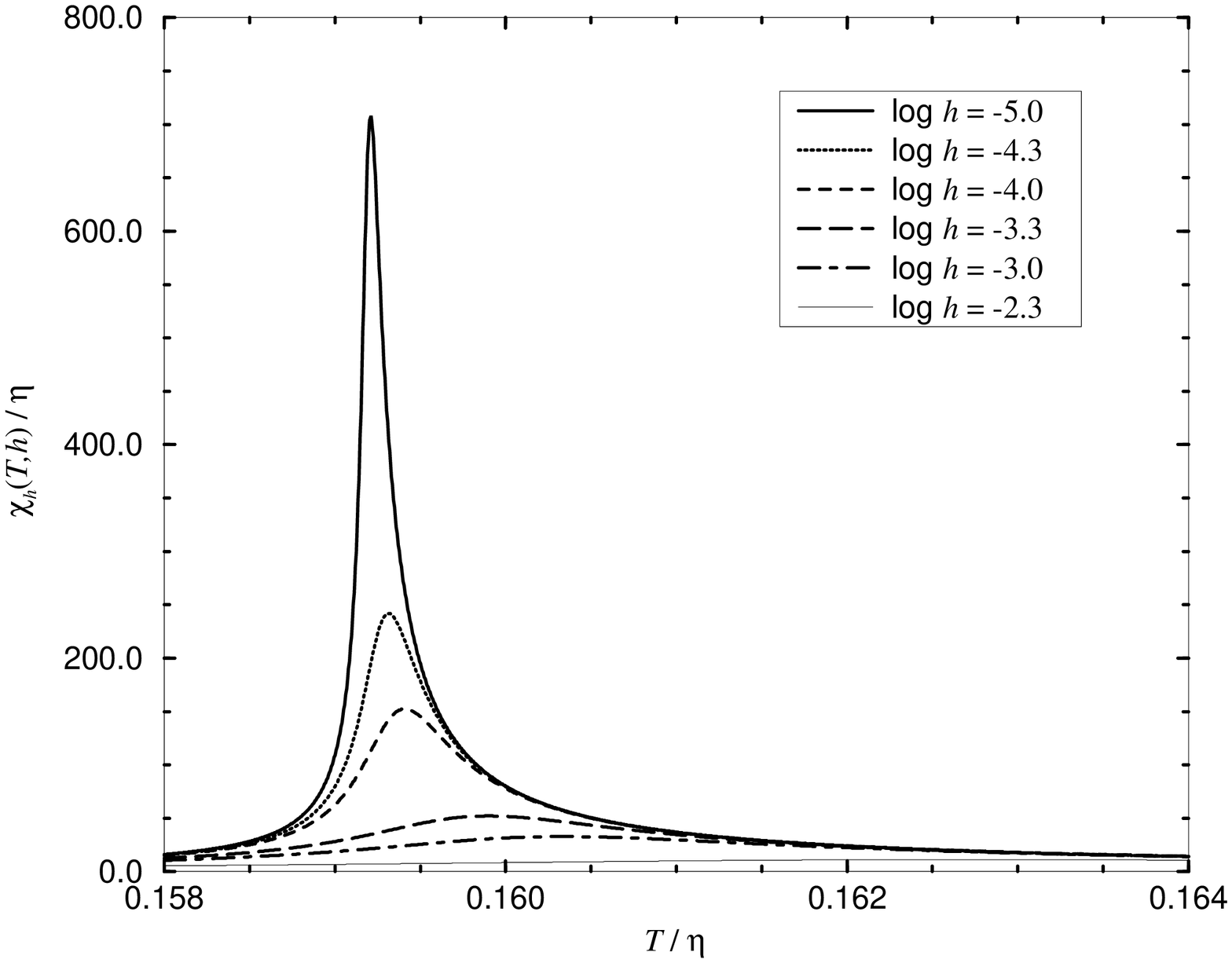,height=12.0cm,angle=-90}}
\caption{The chiral susceptibility, $\chi_h(T,h)$, in the infrared-dominant
model, Eq.~(\protect\ref{chihmn}), as a function of $T$ for various values of
$h$.
\label{mncs}}
\end{figure}

\begin{figure}[h]
 \centering{\
 \epsfig{figure=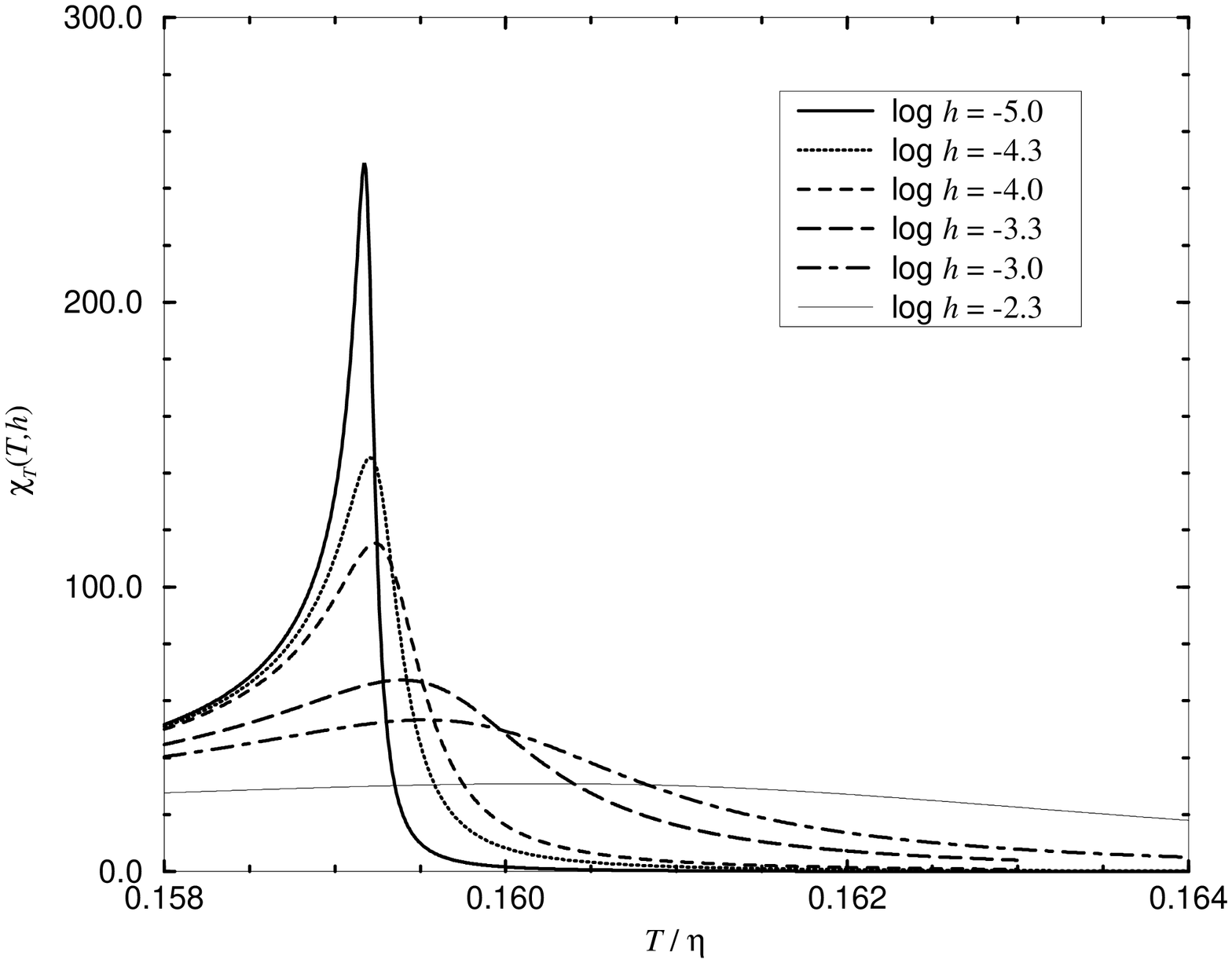,height=12.0cm,angle=-90}}
\caption{The thermal susceptibility, $\chi_T(T,h)$, in the infrared-dominant
model, Eq.~(\protect\ref{chiTmn}), as a function of $T$ for various values of
$h$.
\label{mnts}}
\end{figure}
\begin{figure}[h]
 \centering{\
 \epsfig{figure=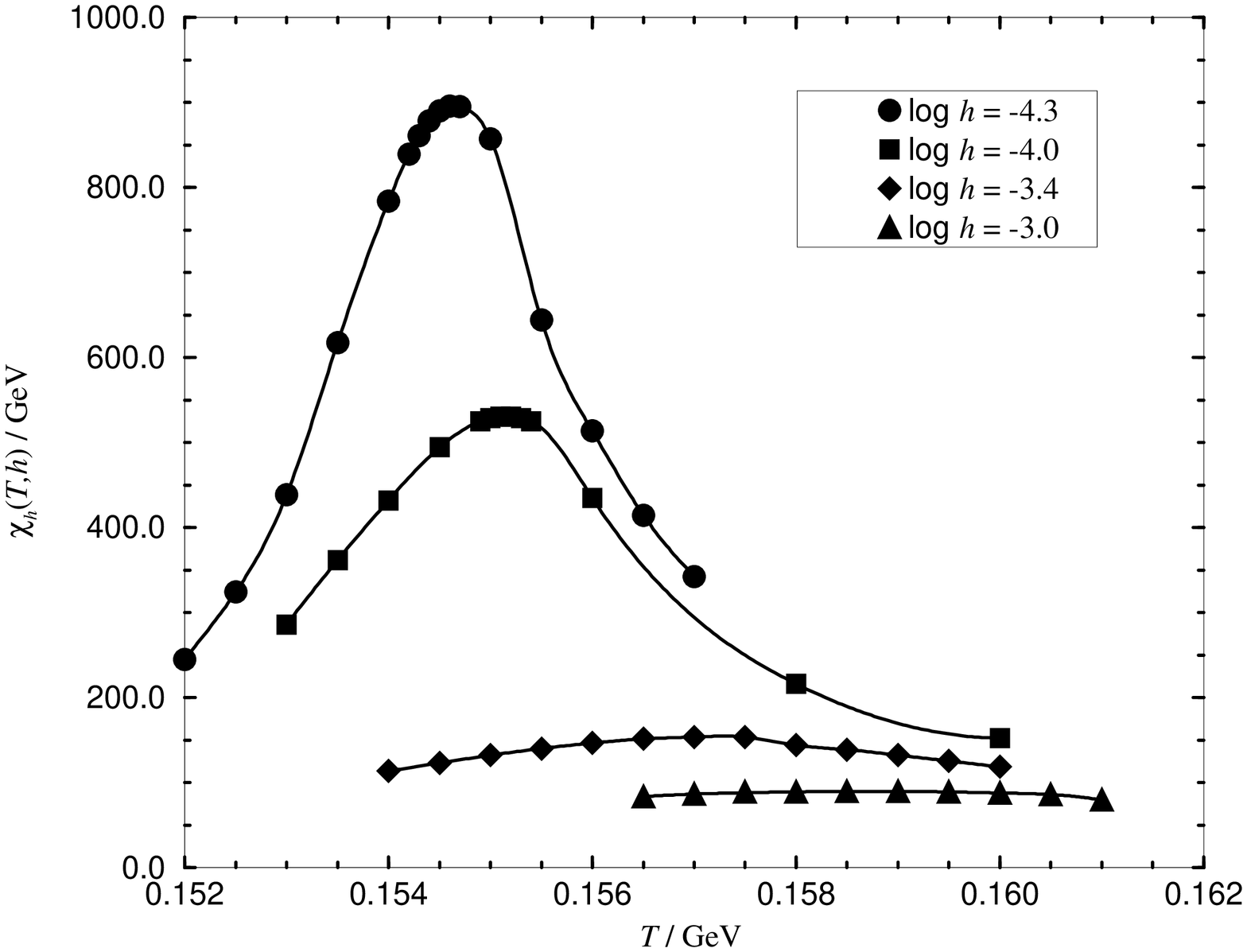,height=12.0cm,angle=-90}}
\caption{The chiral susceptibility, $\chi_h(T,h)$, in the
ultraviolet-improved model, Sec.~\protect\ref{secuvi}, as a function of $T$
for various values of $h$.
\label{frcs}}
\end{figure}

\begin{figure}[h]
 \centering{\
 \epsfig{figure=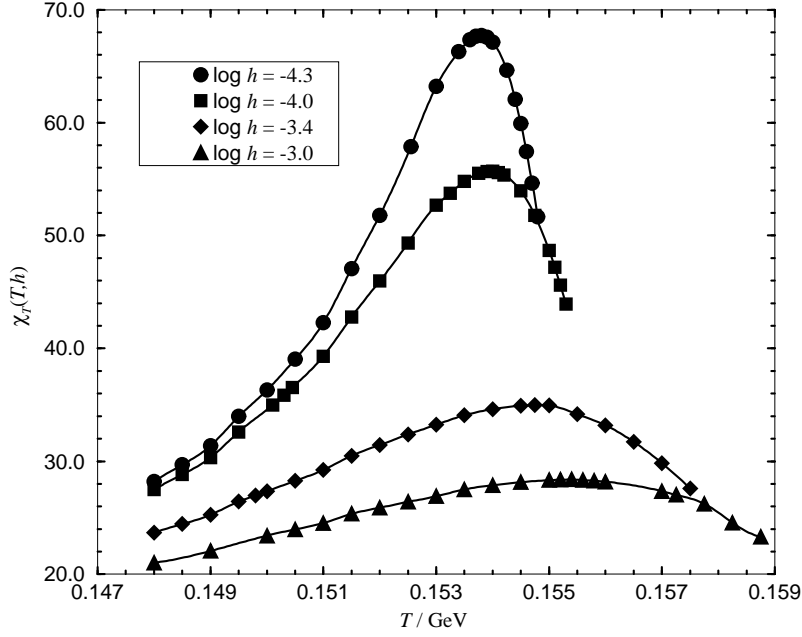,height=12.0cm,angle=-90}}
\caption{The thermal susceptibility, $\chi_T(T,h)$, in the
ultraviolet-improved model, Sec.~\protect\ref{secuvi}, as a function of $T$
for various values of $h$.
\label{frts}}
\end{figure}

\begin{figure}[h]
 \centering{\
 \epsfig{figure=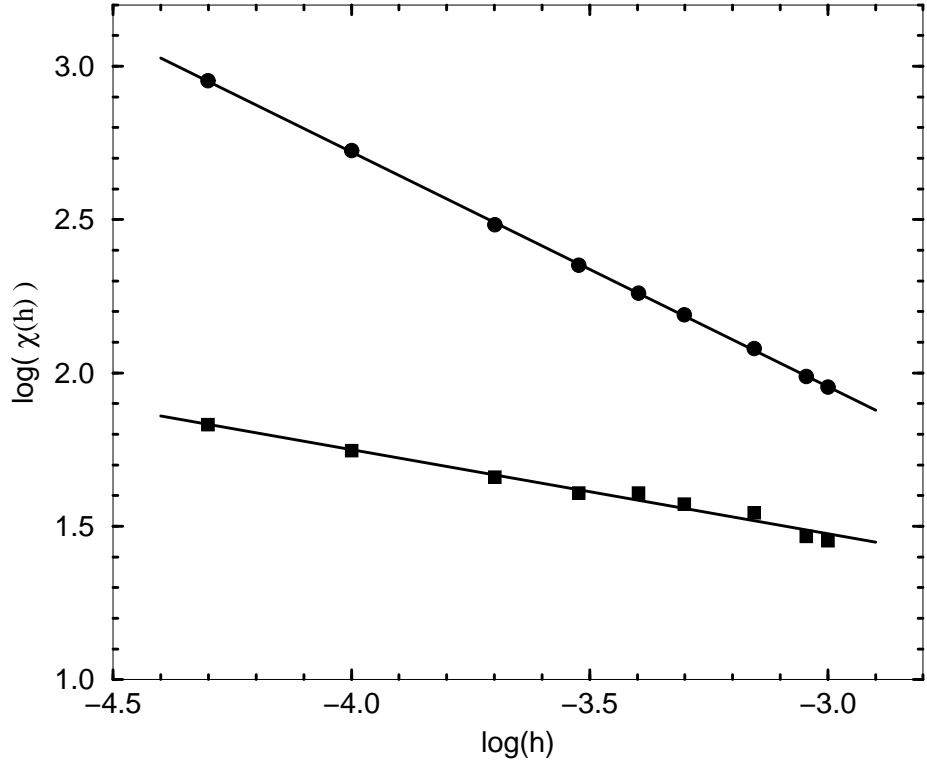,height=12.0cm}}
\caption{The peak heights at the pseudocritical points of the chiral and
thermal susceptibilities in the ultraviolet-improved model: $\chi_h^{\rm pc}$
(filled-circles), $\chi_T^{\rm pc}$ (filled squares).  The solid lines are
straight-line fits, with the slopes $-z_h^{\rm UV}$ and $-z_t^{\rm UV}$ given
in Eq.~(\protect\ref{zeds}), which verify the scaling laws in
Eqs.~(\protect\ref{deltaslope}) and (\protect\ref{betaslope}).
\label{figchis}}
\end{figure}

\end{document}